\documentclass[options]{JHEP3}

\title{M2-branes wrapped on holomorphic curves}

\author{Tasneem Zehra Husain\\ Department of Physics,\\ 
Stockholm University, \\PO Box 6730,\\
S 11385 Stockholm,\\
Sweden.\\ 
Email: \email{tasneem@physto.se}}

\abstract{The generalised calibration for a wrapped membrane is gauge 
equivalent to the supergravity three-form under which the membrane is 
electrically charged. Given the relevant calibration, one can go a long way towards 
constructing the supergravity solution for the wrapped brane.
Applications of this method have been restricted since  
generalised calibrations have not yet been completely classified in spacetimes
with non-vanishing flux. In this paper, we take a first step towards such a classification by 
studying membranes wrapping holomorphic curves. Supersymmetry preservation 
imposes a constraint on the Hermitean metric in the embedding space and it is found 
that this can be expressed as a restriction 
on possible generalised calibrations. Allowed calibrations in a particular spacetime are 
simply those which satisfy the constraint equation relevant to that background; in particular, 
we see that the previously considered Kahler calibrations are just a subclass of possible solutions.}


\preprint{USITP-02-03\\ hep-th/0211030}

\newcommand{\no}{{\noindent}}

\newcommand{\C}{{\mathbb C}}

\newcommand{\comment}[1]{}

\def\bbbz{{\sf Z\!\!\!Z}}
\def\sl2z{SL(2,\bbbz)}
\newcommand{\be}{\begin{equation}}
\newcommand{\ee}{\end{equation}}
\newcommand{\bea}{\begin{eqnarray}}
\newcommand{\eea}{\end{eqnarray}}

\def\bbbz{{\sf Z\!\!\!Z}}
\def\sl2z{SL(2,\bbbz)}

\def\z0{{\bf z_0}}

\newcommand{\bit}{\begin{itemize}}
\newcommand{\eit}{\end{itemize}}

\begin{document}

\setcounter{page}{1}
\pagestyle{plain}

\section{Introduction}

Solitonic M2-branes\footnote{A parallel discussion can be carried out for M5-branes, 
and will appear in \cite{M5}} are like building blocks for the BPS spectrum 
of M-Theory. Flat membranes preserve half of the space-time supersymmetry and a 
large number of supersymmetric states, preserving varying fractions of supersymmetry,
 can be generated either by wrapping membranes 
on holomorphic cycles, or by considering BPS configurations of intersecting membranes. 
Classical solutions corresponding to these systems exist in 11-dimensional supergravity and 
have been studied for a wide variety of cases \cite{wrappedmem}.

We start in Section \ref{mem} by reminding ourselves of some facts about 
membranes which will be used throughout the paper. In Section \ref{Cal} 
we introduce the notion of a calibration \cite{HarveyLawson} and then  
 extend it to define a generalised calibration \cite{GenCal} 
(see also \cite{AdS}). For detailed discussions of these concepts, refer to 
\cite{LecNotes}.

In this paper, we consider only those static supersymmetric solutions of
11-dimensional supergravity which describe membranes wrapping holomorphic curves. 
Supergravity solutions for such branes can be found in several ways, of which
we focus on two, in Section \ref{SugraSoln}. Both these methods require us to 
postulate a form for the metric, so in section \ref{metric}, following Fayyazuddin and Smith we write down 
an ansatz which captures the isometries of the brane configuration.   

In Section \ref{FSmethod}, we 
construct the supergravity solution by looking 
for bosonic backgrounds which admit Killing spinors \cite{FS}. In addition to 
the four-form and functions in the metric ansatz, our analysis yields a 
constraint on the metric in the complex 
subspace where the supersymmetric cycle is embedded.

In Section \ref{SG2}, we adopt an alternate route \cite{Amherst} to solving the 
same problem using the technology of (generalised) calibrations. This procedure is 
simpler by far, but it can only be applied 
if we are given a generalised calibration to start with. Since there is not yet an 
exhaustive list to choose from, applications of this method were restricted to 
Kahler calibrations only. Finally, the labour of section \ref{FSmethod} pays off; 
the metric constraint we obtained is reformulated to  
determine a class of generalised calibrations which exists in backgrounds with 
non-vanishing four form flux. This extends the class of spacetimes in which the 
method of \cite{Amherst} can be used, to find supergravity solutions. 

In his talk at Strings 2000, I remember Sunil Mukhi motivating some examples
with the justification that {\it ``.. You wrap the branes on everything possible, 
until you're blue in the face''}. Even though we may not 
be able to name exact shades yet, this paper attempts to at least figure out
the basic colours your face might turn before you're done with wrapping membranes
on holomorphic curves in $\C^n$! 

\section{Membranes: A Cheat Sheet}
\label{mem} 

We start by collecting a few facts, both to serve as a quick reminder and to set 
notation for what follows. For simplicity, we consider a flat membrane with world-volume 
$012$; $\mu = 0,1,2$ denotes directions along the membrane, and $\alpha = 3,4, ..., 10$ spans 
directions transverse to the membrane.\\

\no
\underline{Killing Spinors}: The spacetime supersymmetries preserved by the M2-brane
correspond to the 16 components of the Killing spinor 
$\chi$ which obeys the projection ${\hat \Gamma}_{012} \chi = \chi$.\\

\no
\underline{The Supergravity Solution}: The bosonic fields in the supergravity solution are the metric and 
a supergravity three form which couples electrically to the membrane.
For the flat membrane described above, the metric takes the form:
\be
ds^2 = H^{-2/3} \eta_{\mu \nu} dX^{\mu} dX^{\nu} + 
H^{1/3} \delta_{\alpha \beta} dX^{\alpha} dX^{\beta}
\ee
and the field strength of the three form is given by: 
\be
F_{012\alpha} = \frac{1}{2} \frac{\partial_{\alpha} H}{H^2} 
\ee
where $H= 1 + \frac{a}{r^6}$ is a harmonic function of the transverse radial 
coordinate $r$.\\

\no
\underline{Wrapped and Intersecting Membranes}: Intersecting branes 
(of the same type) can often be obtained as the singular limit of 
branes wrapping smooth cycles. The simplest and most frequently quoted example 
is that of a single membrane wrapped on the holomorphic 
curve $f(u,v) = uv - c = 0$ in $\C^2$, where $c$ is a constant.  
In the limiting case when $c$ = 0 this curve becomes singular and can 
be described as a system of two orthogonal membranes which span the $u$ and $v$-planes
respectively and intersect at a point. In general, a complex structure can be defined 
on the relative transverse 
directions\footnote {Relative transverse directions are defined to be 
those which are common to at least one but not all of the 
constituents of a system of intersecting branes} and the intersecting brane 
configuration describes the singular limit of a membrane wrapping a 
holomorphic cycle in this complex subspace\footnote{The $(p-2)$ self intersection rule, (which arises from the 
demand that brane intersections be dynamical objects in the world-volume theory),  
states that BPS configurations can be constructed from a number of flat membranes
if these are placed \cite{prule} such that no spatial direction is shared by any pair; 
a supersymmetric system of $n$ intersecting membranes thus has a $2n$-dimensional 
relative transverse space.}.  

We do not 
expect the amount of supersymmetry to vary as we change the constant in 
the holomorphic function, so the Killing spinors of the wrapped brane 
configuration ($c \neq 0$) should be the same as those for a system of $n$ 
orthogonal non-overlapping membranes (the $c = 0$ limit). While we will be considering 
only wrapped branes, the intersecting brane limit serves as a useful check when 
figuring out the amount of supersymmetry preserved by the configuration. Since each additional orthogonal
membrane cuts SUSY down by a factor of 2, a system of $n$ membranes preserves $1/2^n$ of the total supersymmetry;
this then, should be the amount of supersymmetry preserved by a single brane wrapping a holomorphic two-cycle in 
$\C^n$.

\section{Calibrations: Standard and Generalised}
\label{Cal}

Calibrations $\phi_p$ are $p$-forms, characteristic of a particular space-time, which enable us
to classify the minimal $p$-dimensional submanifolds which exist in
that background. A $p$-form $\phi_p$ is a standard calibration if 
\bea
d \phi_p &=& 0\\
| {\cal P} (\phi_p) |\; &\leq& \; | dV_{{\cal M}_p}|
\label{caldefn}
\eea
A manifold $\Sigma_p$ which saturates the above inequality is known as a calibrated manifold.

\no
For backgrounds with no flux, supersymmetry preservation requires the 
existence of covariantly constant spinors on the compactification manifold. 
These in turn imply that the manifold has reduced holonomy;  
calibrations on such manifolds have been classified (see for instance 
\cite{GPLWCal}) and include Kahler and Special Lagrangian calibrations. 
Since in the absence of flux only minimal volume branes are stable, it follows 
that the volume form on a stable brane must be the pullback of a calibrated form in 
the ambient spacetime!\\ 

\no
\underline{Generalised Calibrations}: Given that 
calibrations emerge so naturally in the context of BPS branes, it is tempting 
to try and extend the concept of calibrations to include a treatment of charged $p$-branes.
This turns out to be non-trivial; a charge gives rise to a field strength, 
contradicting the no-flux assumption which lead to all the earlier simplifications.
Taking this into account, generalised calibrations $\phi_p$ are defined \cite{GenCal} such that  
\bea
d (A_{p} + \phi_p) = 0 \;\;\;\;\;\;\; &\Rightarrow& \;\;\;\;\;\;\; 
F_{(p + 1)} = d \phi_p \\
{\rm and} \;\;\;\;\;\;\; |{\cal P} (\phi_p)| 
\; &\leq& |dV_{\Sigma_p}| 
\label{gencaldefn}
\eea
hold, for any $p$-dimensional submanifold ${\Sigma_p}$. 

The most important difference 
between calibrations and generalised calibrations is 
that for the latter, the forms $\phi_p$ are no longer closed\footnote{The results of the previous section
are recovered, as $d \phi = 0$ 
when $F = 0$.}. Notice that the invariant volume form $dV_{\Sigma_p}$
carries a non-trivial contribution from the determinant of the metric on the brane. 

For a $p$-brane wrapped on an $m$-cycle, $l + 1$ worldvolume directions
remain unwrapped ($p = l + m$). The electric potential $A_{p+1}$ to which the brane couples, 
is gauge equivalent to its generalised calibration. This generalised calibration however, lives in 
the full spacetime and we are mostly interested in the embedding space; a generalised calibration 
$\phi_m$ can be defined on this subspace, through \cite{GenCal}
\be
A_{p+1} = dV_{l+1} \wedge \phi_m
\label{gcaldef}
\ee 
where $dV_{l+1}$ is the curved space volume form in the $(l+1)$ unwrapped directions. We will make use of this
definition later, when we discuss membranes wrapped on holomorphic curves and want to 
focus on generalised calibrations in the complex subspace.  

\section{Membranes wrapped on holomorphic curves $\Sigma$ in $\C^n$.}
\label{SugraSoln}

In this section, we will make use of two different methods to 
find the supergravity solutions for M2-branes wrapping holomorphic curves 
in $\C^n$, for $n = 2, .., 5$; $n=2$ corresponding to the smallest complex 
manifold in which a two-cycle can be non-trivially embedded, and $n=5$ being  
the largest complex manifold that can be contained in 11-dimensional spacetime.

As mentioned earlier, the common starting point for both these methods is an ansatz for
the space-time metric, so we present this now before we go any further. 

\subsection{Metric Ansatz}
\label{metric}

If it is to describe a static brane, all functions in
the metric must be independent of time. Furthermore, for a membrane 
wrapped on a two-cycle in $\C^n$, rotational symmetry should be preserved in the 
$(10 - 2n)$ spatial directions transverse to the brane. Based on these expected isometries, the
Fayyazuddin-Smith \cite{FS} ansatz for a metric describing the supergravity 
background created by an M2brane wrapping a holomorphic curve in $\C^n$ is\footnote
{This ansatz was used in \cite{Amherst} subject to the restriction that
the metric in the complex subspace is Kahler. We are not making that assumption here}:
\be
ds^2 = - H_1^2  dt^{2} + 2 H_1^{-1} g_{M {\bar N}} dz^{M}
dz^{\bar N} + H_2^2 \delta_{\alpha \beta} dX^{\alpha} dX^{\beta}.
\label{eq:standard}
\ee
Here $z^{M}$ are used to denote the $n$ complex coordinates, 
$X^{\alpha}$ span the remaining $(10 - 2n)$ transverse directions and the Hermitean metric 
in the complex space has been rescaled for latter convenience. We demand that (\ref{eq:standard})
describes a supersymmetric configuration, and thus satisfies 
\be
\delta_{\chi}\Psi_{I} = 
(\partial_{I}  + \frac{1}{4} \omega_I^{ij} \hat{\Gamma}_{ij}
+ \frac{1}{144}{\Gamma_{I}}^{JKLM}F_{JKLM} 
-\frac{1}{18}\Gamma^{JKL}F_{IJKL})\chi 
\label{susy} = 0.
\ee

\subsection{Supergravity Solutions}

\label{FSmethod}

Following the method employed in \cite{FS}, we find bosonic solutions 
to 11-dimensional
supergravity by looking for backgrounds which admit Killing spinors. 
Having set the gravitino to zero, we have made sure that the supersymmetric 
variations of the bosonic fields vanish identically and it is left only to 
impose that the supersymmetry 
variation of the gravitino vanish as well. 

We require that this be true for our metric ansatz (\ref{eq:standard}), 
when the variation parameter in the supersymmetry transformation is a Killing 
spinor. If the metric and four-form thus obtained satisfy the Bianchi Identity and 
the equations of motion for the field strength, they are guaranteed to obey 
Einstein's equations also, thereby furnishing a bosonic solution to 11-dimensional supergravity.   

The first step in this process is to calculate the Killing spinors, and this is what we proceed to do now.

\subsubsection{Killing Spinors}

A $p$-brane placed in a flat space-time, deforms the surrounding 
geometry. Supersymmetries preserved by the newly curved background can be found
using the 'probe brane' approach whereby we introduce another $p$-brane, 
parallel to the one which caused the geometry to deform and calculate 
its Killing spinors. We call this second brane a probe because its effect on the 
geometry is ignored; this does not lead to any problems since 
the supersymmetry preserved is independent of the back-reaction 
of the probe\cite{BBS}. The supersymmetry preserved by a $p$-brane with 
worldvolume $X^{M_0}...X^{M_p}$ is given by the number of spinors which
satisfy the equation 
\be
\chi = \frac{1}{(p+1)!} \epsilon^{\alpha_0 ... \alpha_p}
\Gamma_{M_0 ... M_p} 
\partial_{\alpha_0} X^{M_0} ....
\partial_{\alpha_p} X^{M_p} \chi
\ee
where $\Gamma_{M_0 ... M_p}$ denotes the anti-symmetrized product
of $(p+1)$ eleven-dimensional $\Gamma$ matrices. The Killing spinors 
of a membrane wrapping a holomorphic curve $\Sigma$ in $\C^n$ are then given by:
\be
\Gamma_{0 m \bar{n}} (
\partial_{1} X^{m} \partial_{2} X^{\bar{n}} - 
\partial_{2} X^{m} \partial_{1} X^{\bar{n}}) \chi = 
\sqrt{{\rm det} \; h_{ab}} \; 
\chi
\ee
where 
\be
h_{ab} = ( \partial_{a} X^{m} \partial_{b} X^{\bar{n}}
+ \partial_{b} X^{m} \partial_{a} X^{\bar{n}} ) \eta_{m \bar{n}}, 
\ee
is the induced metric on $\Sigma$. Using the fact that $\chi$ is 
Majorana and thus of the form\footnote{C denotes
the charge conjugation matrix}
\be
\chi = \alpha + \beta = \alpha + C \alpha^{*},
\label{eq:chi}
\ee
we can express the constraints on the Killing spinors as:
\bea
\Gamma_{m \bar{n}} \alpha &=& - \eta_{m \bar{n}} \alpha \\ 
\Gamma_{m \bar{n}} \beta &=& \; \eta_{m \bar{n}} \beta \\ 
i \; \Gamma_{0} \alpha &=& \; \alpha \\
i \; \Gamma_{0} \beta &=& - \beta
\label{eq:KS}
\eea
A membrane wrapped on a holomorphic curve in $\C^{n}$ preserves $\frac{1}{2^{n}}$ 
of the spacetime supersymmetry, corresponding to the $2^{(5 - n)}$ spinors which
satisfy the above conditions.

The flat space Clifford algebra written in complex coordinates\footnote
{$\Gamma$ matrices for the complex coordinates are defined as 
$\Gamma_{z} = \frac{1}{2} (\Gamma_x + i \Gamma_y)$ and 
$\Gamma_{\bar{z}} = \frac{1}{2} (\Gamma_x - i \Gamma_y)$.} takes the form 
\be
\{ \Gamma_{m}, \Gamma_{{\bar{n}}} \} = 2 \eta_{m {\bar{n}} } 
\ee
which resembles the algebra of fermionic creation and annihilation operators.
A spinor $\chi$ in $\C^m$ can consequently be expressed as a sum of terms of 
the form $|n_1 ... n_m>$ where $n_i$ denotes the fermionic occupation numbers $(0 
\; {\rm or} \; 1)$ corresponding to the action of the creation operator 
$\Gamma_{z_i}$ acting on the Fock vacuum. Using this construction, the 11-dimensional 
spinors $\alpha$ and $\beta$ in (\ref{eq:chi}) can be decomposed as follows:
\be
\alpha = a \otimes |0 ... 0> \;\;\;\;\;{\rm and }\;\;\;\;\; 
\beta = b \otimes |1 ... 1>
\label{eq:chi2}
\ee
where $a$ and $b$ are spinors in the $(10 - 2n)$ dimensional space transverse to 
$\C^n$ and due to (\ref{eq:KS}), satisfy:
\bea
i \Gamma_{0} \; a &=& \; a \\
i \Gamma_{0} \; b &=& - b
\label{eq:KS2}
\eea

\subsubsection{The Consequences of Imposing $\delta_{\chi}\Psi_{I} =0$}

Having decomposed the Killing spinors as in (\ref{eq:chi2}) we can 
express $\delta \Psi$ as a linear combination of Fock space 
states, using (\ref{susy}). All these states are independent, 
so the coefficients of each are required to vanish seperately, giving 
rise to a set of relations between the metric and four form field strength.

Since the supergravity solutions for membranes wrapped on holomorphic 
curves in $\C^n$ follow a similar pattern for all $n$, we present the results in 
a unified manner, in order to avoid needless repetition.\\

\no
\underline{Functions in the Metric Ansatz:} The equations that arise from setting the 
gravitino variation to zero allow us to express the functions\footnote{As pointed out in \cite{Amherst}, 
this determines H only up to a rescaling by an arbitrary holomorphic function} in 
the metric ansatz (\ref{eq:standard}) in 
terms of a single function H, as follows
\bea
{\partial}_I \; ln H &\equiv& -3 \; {\partial}_I \; ln H_1^ \\\nonumber
&=& \; 6 \; {\partial}_I ln H_2 \\\nonumber
&=& -{\frac{1}{2}} \; {\partial}_I \; ln ({\rm det} \; g_{M \bar{N}})
\label{eq:harm}
\eea
where $I$ denotes all coordinates in 11-dimensional space-time. This set 
of relations holds for all $\C^n$, with a note of caution to
be sounded for $n=5$; due to the 
absence of transverse directions in this case there is no $H_2$, but H 
continues to be related to $H_1$ and ${\rm det} \; g_{M \bar{N}}$ as 
stated above.\\

\no
\underline{The Field Strength:} Setting $\delta \Psi = 0$ also determines the four-form 
field strength. Non-zero components are
\bea
F_{0 M \bar{N} \alpha} &=& - \frac{i}{2} 
\partial_{\alpha} g_{M \bar{N}}, \\\nonumber
F_{0 M \bar{N} \bar{P}} &=& - \frac{i}{2} 
[\partial_{\bar{P}} g_{M \bar{N}} - 
\partial_{\bar{N}} g_{M \bar{P}} ], \\\nonumber
F_{0 \bar{M} N P} &=& \; \frac{i}{2} 
[\partial_{P} g_{N \bar{M}} - \partial_{N} g_{P \bar{M}}] 
\label{eq:fstrength}
\eea

\no
These expressions hold for all $n$, with an exception for $n=5$; since 
there are no overall transverse directions, the four-form field 
strength can no longer have a $F_{0 M \bar{N} \alpha}$ component. 
The only non-zero contributions in this case come from $F_{0 M \bar{N} \bar{P}}$ 
and its complex conjugate, $F_{0 \bar{M} N P}$, which are still given by 
expressions above.\\

\no
\underline{The Metric Constraint:} The vanishing of the gravitino variation imposes
the constraint:
\be
\partial [H {\omega}^{(n-1)}_g] = 0 
\label{metriccond}
\ee
where ${\omega}_g \equiv i g_{M {\bar N}} dz^M dz^{{\bar N}}$
is the two-form associated with the Hermitean metric. 

Equation (\ref{metriccond}) is the central result of this paper. It enables us to characterize 
complex manifolds in terms of a single, simple condition on their Hermitean forms. Previously, it was assumed
that complex manifolds in the background of membranes wrapping holomorphic curves, were (warped) Kahler. 
We now see that this assumption is not necessary, except in the special case $n=2$. There is in fact, 
a much larger class of complex manifolds at our disposal than we had at first suspected, corresponding to 
the metrics which satisfy the above constraint. 

\subsubsection{The Bianchi Identity \& Equations of Motion}

It is clear, from the expressions (\ref{eq:fstrength}), that 
the gauge potential is given by $A_{0 M \bar{N}} = i g_{M \bar{N}}$. 
Hence $dF = d^2A = 0$ trivially and the Bianchi Identity is satisfied. \\

\no
In order to arrive at the supergravity solution for the wrapped membranes, it is left 
now only to demand that the equations of motion for the field strength also be satisfied. 
These can be written as follows
$${\partial}_I [ \sqrt{h_{11d}} \; F^{IJKL} ] = 0$$ where $h_{11d}$ denotes the determinant of the full eleven 
dimensional metric. For the wrapped membranes under consideration here, we find that 
$\sqrt{h_{11d}} = H^{1/3}$, regardless
of the dimension of the complex subspace. The only non-trivial contribution to the equations of motion comes from 
$${\partial}_I [H^{1/3} \; F^{0M {\bar N} I}] = 0$$ and takes the form of 
a non-linear differential equation 
\be
\partial^2_{\alpha}[H {\omega}^{(n-1)}_g] +
i 2(n-1) \partial {\bar \partial} [H {\omega}^{(n-2)}_g] = 0
\label{eofm}
\ee
where $n$ denotes the dimension of the complex submanifold. 

It is only when this equation has been solved and an explicit expression 
for $g_{M {\bar N}}$ obtained, that the supergravity solution for the 
wrapped membrane can be said to be found. In practise, solving these 
differential equations proves to be a highly non-trivial exercise; one 
which is beyond the scope of this paper. We will content ourselves here with 
expressing all unknown quantities in terms of the Hermitean metric, which can 
in principle be determined from $(\ref{eofm})$. All the information needed 
to specify the supergravity solution is then known.
 
\subsection{The Power of Calibrations}
\label{SG2}

From the discussion of calibrations in section \ref{Cal}, we 
have learnt that the (pullback of the) generalised calibration corresponding 
to a stable wrapped brane is given simply by its volume form. Moreover, 
this calibration is gauge equivalent to the supergravity three form to which 
the membrane couples electrically; the field strength $F= dA = d\phi$ thus 
follows immediately. Supersymmetry requirements fix the undetermined
functions in the metric ansatz in terms of a single function H,  
which is related to the metric through a non-linear differential equation 
(\ref{eofm}) that follows from $d*F=0$. We now illustrate this process\footnote{
Details are given in the original paper \cite{GenCal} and in \cite{Amherst}.} 
by applying it to the case at hand; membranes wrapping holomorphic curves. 

\subsubsection{A Simpler Method}

We will now sketch the steps involved in 
constructing supergravity solutions for holomorphically wrapped membranes in $\C^n$,
using generalised calibrations. 

Starting with the standard metric (\ref{eq:standard}), we can read off the gauge potential 
\be
A =  H_1 \; dt \wedge \; i H_1^{-1} \; g_{M {\bar N}} \; dz^{M} \wedge dz^{\bar N}
\ee
This, in light of the above discussion, is the spacetime three-form which is 
(gauge equivalent to) the calibration for the membrane. Since we would 
like to focus only on the complex manifold and those generalised calibrations which
may exist on it, we 'split up' this three-form into the product of two lower dimensional 
calibrations; a one-form along the time direction, and a two-form in the complex space. 
Comparing the gauge potential above with (\ref{gcaldef}) we find that
\be
\phi_{M {\bar N}} = i H_1^{-1} g_{M {\bar N}} 
\ee
i.e the generalised calibration on the complex subspace is given by the 
Hermitean metric!

In \cite{Amherst}, this procedure was used to construct supergravity solutions for 
wrapped membranes and fivebranes. However, the only calibrations considered there were 
Kahler. This was in no way a limitation of the approach, but merely reflected the absence 
of knowledge regarding other possible calibrations. As we will see in the following, the 
same procedure can be trivially extended to find supergravity solutions for 
a much larger class of calibrated branes.

\subsubsection{Constraints}
\label{mainpoint}

Since the calibration $\phi_{M {\bar N}}$ on the complex subspace is 
so intimately linked to the Hermitean
metric, it is obvious that by restricting the metric to obey a certain
constraint, we are in fact imposing a condition on the generalised calibrations 
which can exist in this space.\\

\no
In terms of the rescaled metric $k_{M \bar{N}} \equiv  H^{1/(n-1)} g_{M \bar{N}}$
and its associated Hermitean form $\omega = i k_{M \bar{N}} dz^{M} \wedge dz^{\bar N}$,
the metric constraint (\ref{metriccond}) is 
\be
\partial \omega^{(n-1)} = 0
\label{omegah}
\ee
Taking this to be the defining relation, we can look for Hermitean forms which satisfy it.
Each such form $\omega$ will give rise to an associated generalised
calibration $\phi$ on the complex space, through
\be
\phi = H_1^{(4-n)/(n-1)} \omega
\ee
It should now be clear that the Kahler calibrations considered in \cite{Amherst} do not
exhaust the possibilities and in fact correspond only to the obvious solutions of (\ref{omegah}) 
for which $\partial \omega = 0$.\\

\no
Alternately, the constraint can be written directly in terms of calibrations. 
It then states that holomorphic two-form 
$\phi_{M \bar{N}}$ is a generalised calibration only if
\be
\partial *_{\C} [ \phi_{M \bar{N}} \sqrt{ {\rm det} h } ] = 0
\label{finalcond}
\ee
where $*_{\C}$ denotes the Hodge dual in the complex subspace and 
$\sqrt{{\rm det} \; h} =  H^{(4 - n)/3}$ 
is the square root of the determinant in the remaining, non-complex, part of space-time.\\

\no
As it was subject to a non-linear differential equation, we did not have an explicit expression
for the Hermitean metric even when it was assumed to be Kahler; we merely expressed the 
four-form and the undetermined functions in our ansatz in terms of the Hermitean metric. 
Note that these expressions remain the same, even for this wider class of solutions; only
the condition on the undetermined Hermitean metric is relaxed. 
 
\section{Summing Up.}
\label{theend}

We have seen that in the presence of a field strength, calibrated manifolds embedded 
in a complex subspace have a non-trivial dependence on the surrounding spacetime 
as well. This can be intuitively understood as follows. When the field strength is zero,
Killing spinors on the complex subspace are covariantly constant with respect to 
the Hermitean metric and the volume of a supersymmetric brane is minimized. 
A non-zero field strength however, generates a flux which curves the background 
geometry. The ${\rm det}\; h$ factor in (\ref{finalcond}) reflects the fact that 
the four-form flux warps the geometry of spacetime, modifying the 
definition of minimal (supersymmetric) cycles in the complex subspace. 
A new metric can be defined which incorporates the effect of the field 
strength into its torsion. With respect to this redefined metric, the  
Killing spinors of the brane configuration will be covariantly constant and, when
measured by the new metric, the world-volume of a supersymmetric brane will be 
minimized.

To reiterate, we have found that membranes yield stable, supersymmetric 
configurations when wrapped on holomorphic
curves in a complex manifold for which some power\footnote{The powers $1$ to $(n-1)$ 
appear explicitly in the constraints above. Higher powers generate forms which are 
trivially closed} of the (rescaled) Hermitean metric is a closed form. The volume 
form of the wrapped membrane is then the pull-back of a generalised calibration in 11-dimensional 
space-time. In particular, the class of $(1,1)$-forms which satisfies (\ref{finalcond}) gives 
a set of generalised calibrations for M2-branes wrapped on holomorphic cycles. Solutions of this condition include, 
but are not restricted to, Kahler calibrations. 

\acknowledgments
I am grateful to Ansar Fayyazuddin for general encouragement and countless discussions 
during the course of this work and to Fawad Hassan for useful comments and a 
thorough reading of the draft. My gratitute also, to the unknown referee whose
helpful criticism improved the quality of this work.\\

\no
{\bf Note:} References \cite{QMW} which appeared after this paper was written also 
discuss the classification of BPS solutions to 11-d supergravity in backgrounds
with non-vanishing flux.

\end{document}